\documentclass[sigconf]{acmart}
\usepackage{xcolor}
\usepackage{placeins}
\AtBeginDocument{%
  }

\renewcommand\footnotetextcopyrightpermission[1]{}
\acmConference{}{}{}
\acmBooktitle{}
\acmPrice{}
\acmDOI{}
\acmISBN{}
\usepackage{adjustbox}
\usepackage{algorithm}
\usepackage{algorithmicx}
\usepackage{algpseudocode}
\usepackage{amsmath}
\usepackage{amsthm}              
\usepackage{array}
\usepackage{balance}
\usepackage{booktabs}
\usepackage[skip=1pt,labelfont=bf]{caption}
\usepackage{calc}
\usepackage{calligra}
\usepackage{color}
\usepackage{colortbl}
\usepackage{courier}
\usepackage{csvsimple}
\usepackage{enumitem}
\usepackage{etoolbox}
\usepackage{fancybox}
\usepackage{fancyvrb}
\usepackage{fontenc}
\usepackage{fontawesome5}
\usepackage{graphicx}
\usepackage{listings}
\usepackage{longtable}
\usepackage{lscape}
\usepackage{makecell}
\usepackage{marvosym}
\usepackage{moreverb}
\usepackage{multicol}
\usepackage{multirow}
\usepackage{pifont}
\usepackage{pgfplots}
\usepackage[figuresright]{rotating}
\usepackage{setspace}
\usepackage{siunitx}
\usepackage{soul}
\usepackage{subcaption}
\usepackage{tablefootnote}
\usepackage[most]{tcolorbox}
\usepackage{threeparttable}
\usepackage{tikz}
\usepackage[normalem]{ulem}
\usepackage{url}
\usepackage{wasysym}
\usepackage{xspace}

\usepgfplotslibrary{statistics}
\usetikzlibrary{matrix, shapes.geometric, arrows}
\usetikzlibrary{shapes, arrows, positioning}

\algnewcommand\algorithmicforeach{\textbf{for each}}
\algdef{S}[FOR]{ForEach}[1]{\algorithmicforeach\ #1\ \algorithmicdo}

\newcolumntype{L}[1]{>{\raggedright\let\newline\\\arraybackslash\hspace{0pt}}m{#1}}
\newcolumntype{C}[1]{>{\centering\let\newline\\\arraybackslash\hspace{0pt}}m{#1}}
\newcolumntype{R}[1]{>{\raggedleft\let\newline\\\arraybackslash\hspace{0pt}}m{#1}}

\definecolor{codegreen}{rgb}{0,0.6,0}
\definecolor{codered}{rgb}{1,0,0}
\definecolor{codegray}{rgb}{0.5,0.5,0.5}
\definecolor{codepurple}{rgb}{0.58,0,0.82}
\definecolor{backcolour}{rgb}{0.95,0.95,0.92}
\definecolor{lightgray}{gray}{0.9}

 { \newcommand{\mynote}[2]{
      \fbox{\bfseries\sffamily\scriptsize#1}
        {\small$\blacktriangleright$\textsf{\emph{#2}}$\blacktriangleleft$}}}
        { \newcommand{\mynote}[2]{}}
        
\definecolor{DarkOrange}{rgb}{0.8,0.3,0.0}
\definecolor{DarkCyan}{rgb}{0.0, 0.55, 0.55}
\definecolor{DarkCyel}{rgb}{1.0, 0.49, 0.0}
\definecolor{yellow-green}{rgb}{0.6, 0.8, 0.2}

\newcolumntype{?}{!{\vrule width 1pt}}

\newcommand{\find}[1]{
\begin{tcolorbox}[leftrule=1mm,toprule=0mm,bottomrule=0mm,left=1pt,right=2pt,top=2pt,bottom=2pt]
#1
\end{tcolorbox}
}

\lstdefinelanguage{mymarkdown}{
    morekeywords={*,\#, \#\#, \#\#\#},
    sensitive=false,
    morecomment=[l]{//},
    morestring=[b]",
    commentstyle=\color{codegreen},
    keywordstyle=\color{magenta},
    numberstyle=\tiny\color{codegray},
    stringstyle=\color{codepurple},
    basicstyle=\small,
    breakatwhitespace=false,         
    breaklines=true,
    breakindent=0pt,
    keepspaces=true,                 
    numbers=left,                    
    numbersep=5pt,                  
    showspaces=false,                
    showstringspaces=false,
    showtabs=false,                  
    tabsize=2,
}

\lstdefinestyle{mystyle}{
    commentstyle=\color{codegreen},
    keywordstyle=\color{magenta},
    numberstyle=\small\color{black},
    stringstyle=\color{codepurple},
    basicstyle=\scriptsize\ttfamily,
    breakatwhitespace=false,
    breaklines=true,
    captionpos=b,
    keepspaces=true,
    showspaces=false,
    showstringspaces=false,
    showtabs=false,
    tabsize=2
}

\lstdefinelanguage{diff}{
  morecomment=[f][\color{blue}]{@@},     
  morecomment=[f][\color{red}]-,         
  morecomment=[f][\color{codegreen}]+,       
  morecomment=[f][\color{red}]{---}, 
  morecomment=[f][\color{codegreen}]{+++},
  numberstyle=\tiny\color{codegray},
  numbers=left,                    
  numbersep=5pt,         
}

\setlist{noitemsep} 

\definecolor{darkpastelred}{rgb}{0.76, 0.23, 0.13}
\definecolor{ao(english)}{rgb}{0.0, 0.5, 0.0}

\definecolor{darkpastelred}{rgb}{0.76, 0.23, 0.13}
\definecolor{ao(english)}{rgb}{0.0, 0.5, 0.0}

\newboolean{useblue}
\setboolean{useblue}{false} 

\newcommand{\maybeblue}[1]{%
    \ifthenelse{\boolean{useblue}}%
    {\textcolor{blue}{#1}}%
    {#1}%
}

\colorlet{rqaccent}{blue!60}
\colorlet{rqframe}{blue!35}
\colorlet{rqback}{blue!3}

\newtcolorbox{rqbox}[1]{enhanced,breakable,
  colback=rqback, colframe=rqframe,
  boxrule=0.5pt, arc=1.2mm,                 
  borderline west={2pt}{0pt}{rqaccent},     
  left=6pt,right=6pt,top=6pt,bottom=6pt,
  fonttitle=\bfseries, title={Answer to #1},
  before skip=6pt, after skip=6pt}
\begin{document}
\newboolean{showcomments}
\setboolean{showcomments}{true}

\definecolor{DarkOrange}{rgb}{0.8,0.3,0.0}
\definecolor{DarkCyan}{rgb}{0.0, 0.55, 0.55}
\definecolor{DarkCyel}{rgb}{1.0, 0.49, 0.0}
\definecolor{yellow-green}{rgb}{0.6, 0.8, 0.2}
\definecolor{DarkGreen}{RGB}{0,128,0}
\newcommand{\up}[1]{\textcolor{DarkGreen}{(+#1\%)}}
\newcommand{\down}[1]{\textcolor{red}{(-#1\%)}}
\newcommand{\same}{\textcolor{gray}{(0.0\%)}}

\newcommand{\todoc}[2]{{\textcolor{#1} {\textbf{#2}}}}
\newcommand{\todoblue}[1]{\todoc{blue}{\textbf{#1}}}
\newcommand{\todogreen}[1]{\todoc{yellow-green}{\textbf{#1}}}
\newcommand{\todored}[1]{\todoc{red}{\textbf{#1}}}

\newcommand{\yang}[1]{\mynote{Boyang}{\todogreen{#1}}}
\newcommand{\zijian}[1]{\mynote{Zijian}{\todoblue{#1}}}

\newcommand{\tool}{\textsc{PAFT}\xspace}
\newcommand{\baseline}{\textsc{Base}\xspace}
\newcommand{\stdft}{\textsc{Sft}\xspace}
\newcommand{\dataset}{\textsc{TutorLLMCode}\xspace}
\newcommand{\prompt}{\textsc{Prompting}\xspace}

\title{\tool: \underline{P}reservation-\underline{A}ware \underline{F}ine-\underline{T}uning for Minimal-Edit Program Repair}
\author{Boyang Yang}
\affiliation{%
  \institution{School of Artificial Intelligence (School of Software), Yanshan University}
  \city{Qinhuangdao}
  \country{China}
}
\email{yby@ieee.org}

\author{Zijian Cai}
\affiliation{%
  \institution{School of Artificial Intelligence (School of Software), Yanshan University}
  \city{Qinhuangdao}
  \country{China}
}
\email{zijiancai6@gmail.com}

\author{Shunfu Jin}
\affiliation{%
  \institution{School of Artificial Intelligence (School of Software), Yanshan University}
  \city{Qinhuangdao}
  \country{China}
}
\email{jsf@ysu.edu.cn}

\author{Haoye Tian}
\authornote{Corresponding author.}
\affiliation{%
  \institution{Department of Computer Science, Aalto University}
  \city{Espoo}
  \country{Finland}
}
\email{tianhaoyemail@gmail.com}

\begin{abstract}
Large language models (LLMs) are effective for automated program repair, but plausible patches that pass the full test suite often rewrite more code than necessary, increasing review and maintenance costs. This over-editing is common because most bugs are localized, while standard supervised fine-tuning provides no explicit signal about which tokens should be preserved and which should be changed. We propose \tool, a preservation-aware fine-tuning method for minimal-edit program repair. \tool derives token-level preservation signals by aligning buggy and fixed code, combines them with full-sequence masking, and applies an edit-difficulty curriculum. Across Defects4J and HumanEval-Java, \tool improves pass@1 by up to 65.6\% over standard supervised fine-tuning (\stdft) while reducing average edit distance (AED) by up to 32.6\%. On Defects4J with DeepSeek-Coder-6.7B, \tool also outperforms AdaPatcher, a strong preference-based repair baseline, improving pass@1 from 5.9\% to 10.1\% while reducing median AED from 61.0 to 42.0. Overall, \tool preserves stable context and concentrates edits on faulty regions, yielding smaller, more localized, plausible patches without inference-time search, reranking, or post-processing.
\end{abstract}



\maketitle

\section{Introduction}

Large language models (LLMs) have become a dominant foundation for automated program repair (APR)~\cite{le2019automated,xia2023automated,yang2025survey,jiang2023impact}. In practice, however, a useful repair patch is not merely one that passes the available tests. It should also keep edits localized to the fault-relevant region and preserve stable surrounding implementation whenever broader rewriting is unnecessary. This matters because patches are reviewed, revised, and merged under maintenance constraints, and code review studies have shown that change size, change decomposition, and reviewability are closely related~\cite{bacchelli2013expectations,10.1145/3236024.3236080,di2019effects}. Yet recent evidence shows that LLM-generated patches often exceed human reference patches in size, indicating systematic over-editing~\cite{yang2024cref}. Over-editing is especially problematic in APR because most tokens in the buggy input belong to a stable context, while only a small region typically requires modification. Within the broader goal of maintainable patch generation, this paper focuses on one central bottleneck: preventing unnecessary rewriting of stable code. Figure~\ref{fig:paft-example} illustrates this failure mode and the minimal-edit behavior we target.

A key reason this problem persists is that current LLM-based APR is still largely realized through prompting or standard supervised fine-tuning~\cite{xia2023automated,10.1109/ASE56229.2023.00181}. Recent work has further shown that the way repair is aligned with the model's underlying objective materially affects APR performance~\cite{xu2025aligning}. Yet these formulations still optimize generation toward the reference patch without explicitly distinguishing the stable context that should be copied from the fault-relevant code that should be edited. A model can therefore produce a functionally plausible patch while still rewriting surrounding logic that ought to remain intact. Repair correctness and edit locality are only weakly coupled in the training signal, which makes localized repair difficult to learn from standard \stdft alone.

Despite growing interest in more controlled repair, existing work still leaves some gaps. Most APR pipelines are still optimized primarily for test-suite passing~\cite{yang2024cref, xia2024automated, yin2024thinkrepair}. Prompt-based repair and standard fine-tuning can improve repair capability, but they do not make preservation an explicit training target~\cite{xia2023automated,10.1109/ASE56229.2023.00181,xu2025aligning}. Meanwhile, current mechanisms for controlling edit scope are also mostly indirect. Post-hoc reduction methods move minimality control to inference time through repeated validation~\cite{zeller2002simplifying,le2011genprog}; preference-based repair methods instead rely on extra preference construction beyond paired buggy and fixed code, as in AdaPatcher, and connect naturally to DPO- and RLHF-style alignment recipes~\cite{dai2025less,rafailov2023direct,ouyang2022training}.

Another gap concerns evaluation. Existing APR work typically reports pass@k together with patch-size statistics, yet prior APR evaluation research has shown that test-suite metrics are only imperfect proxies for repair reliability and that evaluation choices can materially change conclusions~\cite{DBLP:conf/icse/YiTMBR18,DBLP:journals/jss/LiuLKKLKB21}. For review-oriented repair, the practical question is narrower: among plausible patches, did the model change only what was necessary? Without an automatic edit metric empirically tied to human judgments of patch quality, it remains difficult to compare methods or to optimize minimal-edit behavior in a principled way.

Within the broader goal of maintainable patch generation, this paper studies one measurable subproblem: stable-context preservation under plausible repair, rather than maintainability in full. We ask whether a supervised repair model can be trained to change less code while preserving or improving repair effectiveness, and whether the resulting edit metrics align with human judgments of review-oriented patch quality.

We address these gaps with \tool, a preservation-aware fine-tuning framework for minimal-edit program repair. \tool starts from a simple observation: paired buggy and fixed code already contains a usable preservation signal. We align buggy and fixed code with a deterministic token-matching procedure, derive token-level preservation signals, and upweight aligned target tokens during fine-tuning so that the model learns to copy stable spans and concentrate edits on the faulty region. We further combine this objective with full-context supervision and an edit-difficulty curriculum to stabilize training. This design moves minimality control into training itself, without extra preference labels, inference-time reduction, reranking, or search.

Across Defects4J and HumanEval-Java, \tool improves pass@1 by up to 65.6\% over standard \stdft and reduces average edit distance (AED) by up to 32.6\%. On Defects4J, it also outperforms representative prompt-based, fine-tuning-based, and preference-based baselines. We further show, through targeted human annotation, that AED aligns better with review-oriented human judgments than simpler patch-size indicators, while Code Consistency Rate (CCR) provides a complementary view of stable-context preservation.

The main contributions of our work are as follows:

\begin{itemize}[noitemsep,topsep=0pt,leftmargin=2em]
  \item We identify over-editing, namely unnecessary rewriting of stable code, as a distinct failure mode in LLM-based program repair and, through targeted human annotation, show that AED aligns better with review-oriented human judgments than simpler patch-size indicators, with CCR as a complementary preservation metric.
  \item We propose \tool, a lightweight preservation-aware fine-tuning method that derives preservation supervision directly from aligned buggy and fixed code, without extra preference labels, inference-time reduction, reranking, or search.
  \item We conduct comprehensive experiments on Defects4J and HumanEval-Java with three backbone code LLMs, showing that \tool consistently improves pass@1 and, in most settings, improves AED and CCR over standard \stdft. On Defects4J, it also outperforms representative prompt-based, fine-tuning-based, and preference-based baselines.
\end{itemize}

\section{Motivating Example}
\label{app:case-study}

\begin{figure}[h]
  \centering
  \includegraphics[width=\linewidth]{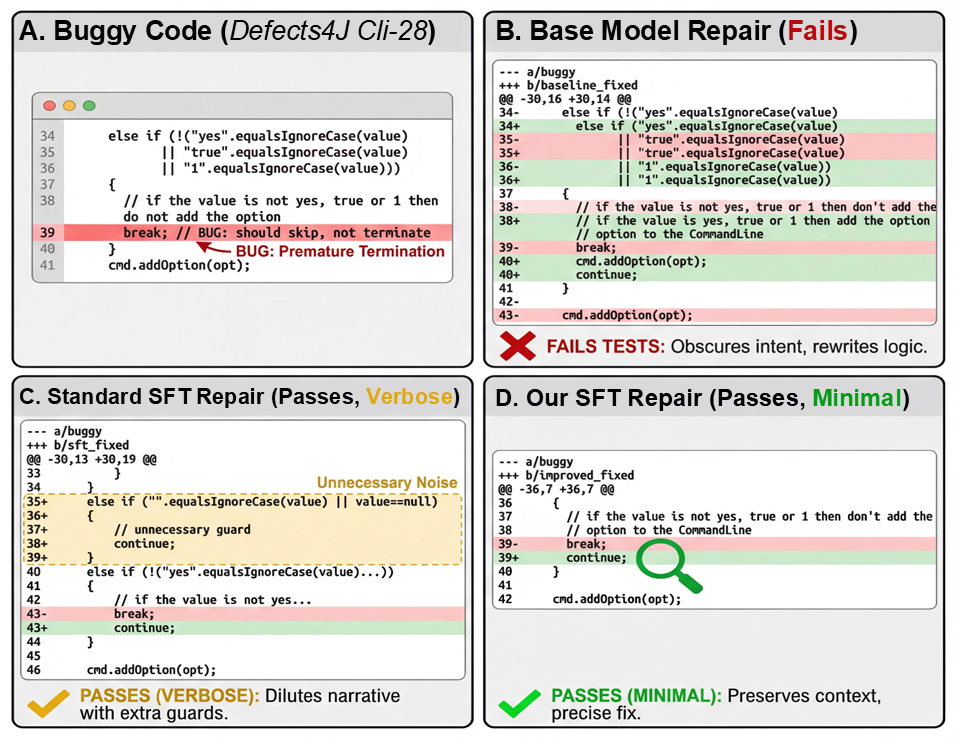}
  \caption{Motivating example of \tool.}
  \label{fig:paft-example}
\end{figure}

Figure~\ref{fig:paft-example} presents a representative case on Defects4J Cli-28 with DeepSeek-Coder-6.7B. This defect is a useful stress test for preservation because the required semantic change is highly localized, while the surrounding context already makes the intended behavior explicit. The loop should process each option independently. When a property value is not affirmative, the correct behavior is to skip only the current option and continue scanning the remaining entries. In the buggy code, however, the loop executes \texttt{break} rather than \texttt{continue}, which terminates the iteration prematurely and prevents later options from being processed. The defect therefore concerns the scope of control transfer, rather than the predicate, the loop structure, or the placement of \texttt{cmd.addOption(opt)}. Because the nearby comment already states the intended behavior, any patch that rewrites substantially more than this control token requires additional justification.

Panel B shows the behavior of the base model. Instead of repairing the control token in place, the model rewrites nearby logic, reformulates the predicate, moves \texttt{cmd.addOption(opt)} into the branch body, and alters the surrounding comment. This produces a much larger diff than the defect warrants and changes the original filter-then-add structure of the loop. The resulting patch is superficially plausible, but it mixes the true fix with predicate rewriting and statement motion; in this case, the extra structural drift also causes the patch to fail tests. This example highlights a common failure mode of unconstrained generation: when the model is not explicitly encouraged to preserve stable spans, it tends to re-express surrounding logic, even when the fault itself is narrow and well-localized.

Panel C illustrates a subtler limitation of standard \stdft. After fine-tuning, the model repairs the premature termination and passes the test suite, but it still adds an explicit guard for \texttt{null} or empty strings that is not required by the defect. Even if the resulting behavior remains acceptable, the patch becomes semantically noisier than necessary. A reviewer must now inspect an additional branch whose connection to the original fault is weak. This is precisely the distinction that pass@$k$ misses: a patch can be plausible and functionally correct, yet still expand the edit scope beyond the causal location of the bug. In review-oriented settings, that extra verification burden matters, because unnecessary control-flow changes make it harder to determine whether the patch is truly faithful to the original implementation intent.

Panel D shows the behavior encouraged by \tool. The model changes only \texttt{break} to \texttt{continue} and preserves the predicate, the comment, and the placement of \texttt{cmd.addOption(opt)} exactly. This one-token repair restores the intended per-option control flow without perturbing stable context. The patch is therefore not only smaller, but also more faithful to the original code narrative, easier to audit, and less likely to introduce incidental behavioral changes outside the true fault. More importantly, it captures the training objective of \tool: copying should be the default behavior on aligned stable spans, while edits should concentrate on the faulty region. The aggregate results in Tables~\ref{tab:defects4j-tool} and~\ref{tab:repairllama-compare} follow the same pattern. Lower AED and higher CCR are therefore not merely indicators of shorter diffs; they reflect whether a model repairs the true fault while preserving the original code structure, a distinction that pass@$k$ alone does not reveal.
\section{Methodology}

\label{sec:approach}

\begin{figure*}[t]
  \centering
  \includegraphics[width=\linewidth]{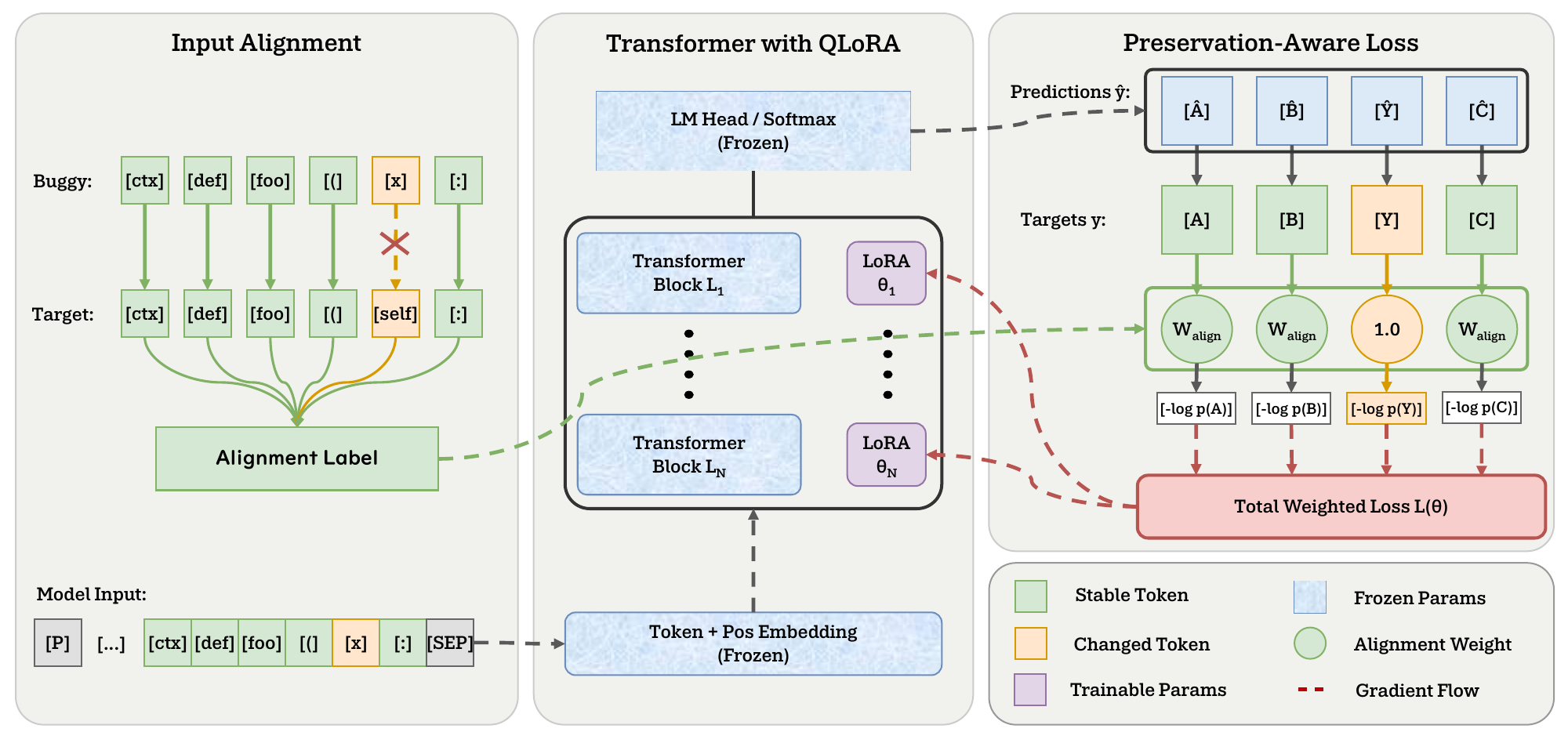}
  \caption{Overview of \tool. We derive an alignment-guided preservation signal from \((p,b,f)\) and fine-tune a quantized code LLM with QLoRA using a preservation-aware token weighting and an edit-difficulty curriculum.}
  \label{fig:paft-overview}
\end{figure*}

\subsection{Data Preparation}
Each training instance is serialized into a single sequence using the backbone's instruction or chat template, where the user segment contains the prompt and buggy context, and the assistant segment contains the reference fix as the target generation. Let \(\mathcal{D}\) be the training corpus built from TutorLLMCode (1535 instances)~\cite{yangmorepair}. For each \((p,b,f)\in\mathcal{D}\), tokenization with the backbone tokenizer yields \(x=(x_1,\ldots,x_{|x|})\).

To derive a preservation signal, we extract the buggy and fixed code segments and tokenize them with the same tokenizer, obtaining token identifier sequences \(T(b)\) and \(T(f)\). We then compute a deterministic token-level alignment between these sequences using a Ratcliff/Obershelp-style pattern matching procedure~\cite{RatcliffMetzener1988Gestalt}. The procedure recursively identifies long contiguous matching spans and yields a set of matching blocks \((i,j,n)\), where the length-\(n\) span starting at position \(i\) in \(T(b)\) matches the span starting at position \(j\) in \(T(f)\). This matching-block alignment is consistent with the widely used \emph{SequenceMatcher} strategy. We treat the union of matched token positions in \(T(f)\) as copy-worthy and map them onto the corresponding response-token positions in \(x\). Let \(\mathcal{I}_{\mathrm{align}}\subseteq\{1,\ldots,|x|\}\) denote the set of response positions whose target tokens fall inside these matched spans. Intuitively, \(\mathcal{I}_{\mathrm{align}}\) identifies response tokens in the fix that can be copied from the buggy context.

Alignment is computed only over code tokens. Specifically, we strip prompt wrappers, role markers, and fenced-block delimiters before matching, and we tokenize the extracted buggy and fixed code with the same backbone tokenizer. After matching, each aligned token position in $T(f)$ is mapped back to its absolute index inside the assistant code span in $x$; tokens outside the extracted fixed-code span keep unit weight. When repeated tokens induce overlapping matching blocks, we take their union in $T(f)$, which yields a deterministic binary preservation mask for the response.

\subsection{Preservation-Aware Fine-Tuning}
We fine-tune 4-bit NF4 quantized backbones with QLoRA adapters~\cite{hu2022lora,dettmers2023qlora}. For an adapted linear layer, let \(W_0 \in \mathbb{R}^{d_{\mathrm{out}} \times d_{\mathrm{in}}}\) denote the pretrained weight matrix. Under QLoRA, the backbone weight is stored in 4-bit NF4 form and dequantized on the forward pass, while a trainable low-rank update is added:
\[
W^{\mathrm{eff}}
\;=\;
\mathrm{Dequant}\!\bigl(\mathcal{Q}_{\mathrm{NF4}}(W_0)\bigr)
\;+\;
\Delta W,
\qquad
\Delta W
\;=\;
\frac{\alpha}{r} B A,
\]
where \(A \in \mathbb{R}^{r \times d_{\mathrm{in}}}\) and \(B \in \mathbb{R}^{d_{\mathrm{out}} \times r}\) are trainable LoRA factors, and \(r\) and \(\alpha\) are the LoRA rank and scaling factor, respectively. For an input hidden state \(h\), the adapted projection is \(h' = W^{\mathrm{eff}} h\). We optimize only the adapter parameters \(\theta = \{A_\ell, B_\ell\}_\ell\) and keep the quantized backbone frozen.

The per-token autoregressive negative log-likelihood is
\[
\ell_t(\theta) \;=\; -\log p_{\theta}\!\left(x_t \mid x_{<t}\right).
\]

We assign a weight to each token position:
\[
w_t \;=\;
\begin{cases}
w_{\mathrm{align}} & \text{if } t \in \mathcal{I}_{\mathrm{align}},\\
1 & \text{otherwise},
\end{cases}
\qquad w_{\mathrm{align}} > 1.
\]

The masked, weighted loss for a training instance is
\[
\mathcal{L}(\theta; x, M)
\;=\;
\frac{\sum_{t=1}^{|x|} M_t\, w_t\, \ell_t(\theta)}
{\sum_{t=1}^{|x|} M_t\, w_t}.
\]

We consider two common types of loss masks: assistant-only masking (loss on response tokens) and full-sequence masking (loss on both prompt and response tokens).
In the main comparisons, \stdft follows the assistant-only convention used in standard instruction tuning, while \tool uses full-sequence masking unless otherwise stated.
Section~\ref{sec:rq4} isolates the impact of masking from preservation-aware token reweighting.
Since \(\mathcal{I}_{\mathrm{align}}\) is defined over response positions, preservation reweighting is applied only to response tokens.
Over the training corpus, we optimize
\[
\theta^\star
\;=\;
\arg\min_{\theta}
\frac{1}{|\mathcal{D}|}
\sum_{(p,b,f)\in\mathcal{D}}
\mathcal{L}(\theta; x, M),
\]
where \(x\) and \(M\) are induced by the instance \((p,b,f)\).

In this paper, \tool denotes the full training recipe consisting of full-sequence masking, an edit-difficulty curriculum, and alignment-based preservation weighting. We retain the assistant-only \stdft baseline because it matches common widely-used instruction-tuning practice, and we isolate the incremental contribution of each added component in Section~\ref{sec:rq4}. We therefore interpret the main RQ-2 results as a comparison between a standard \stdft baseline and the complete \tool recipe, rather than as evidence that token reweighting alone explains the full gain.

\subsection{Curriculum over Edit Difficulty}
To stabilize optimization, we order training examples by an edit-difficulty proxy based on the line-level diff size $\mathrm{dl}(b,f)$.
Let $\ell^{b}=\mathrm{Lines}(\mathcal{N}(b))$ and $\ell^{f}=\mathrm{Lines}(\mathcal{N}(f))$, where $\mathcal{N}(\cdot)$ applies the normalization used for diff computation.
We compute a unified diff between $\ell^{b}$ and $\ell^{f}$, and denote by $a^{\mathrm{line}}(b,f)$ and $d^{\mathrm{line}}(b,f)$ the numbers of added and deleted lines, respectively.
We then define
\[
\mathrm{dl}(b,f) \;=\; a^{\mathrm{line}}(b,f) + d^{\mathrm{line}}(b,f).
\]
Within each epoch, we iterate from smaller $\mathrm{dl}(b,f)$ to larger $\mathrm{dl}(b,f)$, so the model first learns localized edits and then gradually adapts to larger diffs~\cite{bengio2009curriculum}.

\subsection{Repair Inference}
\label{sec:repair-inference}
At inference time, \tool uses the same instruction or chat template as in training to serialize a repair context \(c\), consisting of the prompt and buggy code followed by the assistant-generation prefix. The concrete prompt templates and decoding settings are specified in the experimental setup. Given \(c\), the model generates a repair sequence \(\hat{y} = (\hat{y}_1,\ldots,\hat{y}_m)\) autoregressively:
\[
p_{\theta}(\hat{y} \mid c)
\;=\;
\prod_{t=1}^{m}
p_{\theta}\!\left(\hat{y}_t \mid c, \hat{y}_{<t}\right).
\]

Each conditional distribution is computed by the same QLoRA-adapted model used during fine-tuning. For every adapted layer \(\ell\), the effective inference weight is
\[
W_{\ell}^{\mathrm{eff}}
\;=\;
\mathrm{Dequant}\!\bigl(\mathcal{Q}_{\mathrm{NF4}}(W_{\ell,0})\bigr)
\;+\;
\frac{\alpha}{r} B_{\ell} A_{\ell},
\]
that is, the frozen 4-bit NF4 backbone plus the learned low-rank update. No alignment set \(\mathcal{I}_{\mathrm{align}}\), token weights \(w_t\), or loss mask \(M_t\) is used at inference time.

\tool performs standard left-to-right decoding and produces a candidate patch in one pass, stopping at the end-of-sequence token or a preset generation limit. When multiple candidates are needed for pass@\(k\), we independently repeat the same sampling process under the same model and prompt. \tool does not introduce inference-time alignment computation, edit-distance regularization, reranking, or search. Any behavioral differences relative to \stdft are induced solely by training-time differences in the learned QLoRA adapters, not by changes to the decoding algorithm.
\section{Experimental Setup}
\label{sec:setup}

\subsection{Benchmarks}
\label{sec:setup-benchmarks}

We evaluate \tool on 2 complementary Java repair benchmarks. HumanEval-Java contains function-level buggy programs with unit tests, enabling controlled evaluation within a single-function scope \cite{jiang2023impact}. Defects4J is a project-level benchmark that uses realistic build-and-test harnesses to capture repository-scale defects under compilation and integration constraints \cite{just2014defects4j}. For Defects4J, we directly reuse the benchmark data and evaluation scripts released by MORepair \cite{yangmorepair}. Together, these benchmarks cover both local single-function repair and repository-scale repair under full project constraints.

We follow the same test-driven evaluation workflow as MORepair. A generated patch is counted as plausible only if it compiles and passes the corresponding benchmark test suite. We report pass@$k$ for $k \in \{1,5,10\}$ as the main functional-correctness metric. Beyond test validation, we also evaluate edit magnitude and preservation on plausible patches using the metrics in Section~\ref{sec:metrics}, namely CCR, AED, ATCL, and ATCT.

\subsection{Models}
\label{sec:setup-models}

We evaluate 5 models in total. The 3 open-source backbones that we fine-tune under the unified QLoRA setting are Qwen3-8B \cite{yang2025qwen3}, OpenCoder-8B-Instruct \cite{huang2025opencoder} (denoted as OpenCoder-8B), and Deep\-Seek-Coder-6.7B \cite{guo2024deepseek} (denoted as DS-Coder-6.7B). We additionally report zero-shot reference results from Qwen3-Max \cite{yang2025qwen3} and DeepSeek-V3 \cite{liu2024deepseek} to provide stronger reference points beyond the 6B to 8B fine-tuned setting.

\subsection{Training Data}
\label{sec:setup-data}

For supervised training, we use the TutorLLMCode repair corpus released by MORepair \cite{yangmorepair}. After applying the same parsing rules across all backbones, we obtain 1535 valid buggy-fix pairs in instruction-following format. Each parsed instance contains an \texttt{instruction} field that includes the task statement and buggy program, and a \texttt{response} field that contains the corrected program. From each instance, we extract the prompt $p$, buggy code $b$, and fixed code $f$ for training.

TutorLLMCode also provides an expert-assigned problem-diff\-iculty label on an 11-point scale. Difficulty 1 corresponds to introductory programming exercises, while Difficulty 11 corresponds to top-tier algorithmic contest problems. This label describes the intrinsic difficulty of the underlying problem, not the size of the concrete repair. Table~\ref{tab:dataset-tier} reports the distribution. Overall, 58.18\% of the instances fall into difficulty levels 1--4, whereas only 14.72\% fall into levels 8--11. Therefore the dataset spans a broad pedagogical range, but it is concentrated in the lower-to-middle portion of the scale.

\begin{table}[h]
\centering
\small
\setlength{\tabcolsep}{4pt}
\begin{tabular}{crr|crr}
\toprule
Difficulty & Instances & Share (\%) & Difficulty & Instances & Share (\%) \\
\midrule
1  & 189 & 12.3 & 7  & 155 & 10.1 \\
2  & 184 & 12.0 & 8  & 91  & 5.9 \\
3  & 228 & 14.9 & 9  & 62  & 4.0 \\
4  & 292 & 19.0 & 10 & 67  & 4.4 \\
5  & 72  & 4.7  & 11 & 6   & 0.4 \\
6  & 189 & 12.3 &    &     &     \\
\bottomrule
\end{tabular}
\caption{Expert-annotated problem difficulty of TutorLLMCode (1535 instances).}
\label{tab:dataset-tier}
\end{table}

Our curriculum does not use the expert-assigned problem difficulty above. Instead, it uses an edit-difficulty signal derived directly from each buggy-fix pair. We compute a unified diff between the buggy and repaired code, exclude the diff headers, and count only the added and deleted lines. The resulting diff sizes range from 1 to 119 lines, with a mean of 16.66 and a median of 11. The distribution is right-skewed: 49.25\% of the instances require no more than 10 diff lines, 70.88\% require no more than 20 diff lines, and only 5.34\% involve more than 50 diff lines. Most training examples, therefore, correspond to small-to-moderate local repairs, while a small long-tail subset requires substantially larger edits.

Table~\ref{tab:curriculum_diff_stages} summarizes this edit-difficulty distribution using 3 coarse stages. The table is descriptive rather than operational. During training, we still traverse examples in ascending order by the exact line-diff size $\mathrm{dl}(b,f)$, with ties preserved in the original order. This keeps the curriculum simple, tokenizer-independent, and directly tied to the granularity of edits.

\begin{table}[h]
  \centering
  \small
  \begin{tabular}{lrrrr}
  \toprule
  Stage & Diff-line range & Instances & Ratio (\%) & Mean diff \\
  \midrule
  Stage 1 & 1--5 & 444 & 28.93 & 2.89 \\
  Stage 2 & 6--17 & 563 & 36.68 & 10.51 \\
  Stage 3 & $\geq 18$ & 528 & 34.40 & 34.80 \\
  \midrule
  Total & 1--119 & 1535 & 100.00 & 16.66 \\
  \bottomrule
  \end{tabular}
  \caption{Three-stage summary of the edit-difficulty distribution in the training set, based on diff-line counts.}
  \label{tab:curriculum_diff_stages}
\end{table}

We also check for train-test contamination. We normalize buggy and fixed code snippets with the same function used in our diff metrics, including whitespace normalization and comment removal, and compare SHA-256 hashes between TutorLLMCode and both evaluation benchmarks. We find no exact overlap between any normalized training snippet and any normalized evaluation snippet.

\subsection{Training and Inference Protocol}
\label{sec:setup-protocol}

We fine-tune each open-source backbone with QLoRA under a unified configuration: 4-bit NF4 quantization, maximum sequence length 2048, LoRA rank $r=32$ with $\alpha=16$ and dropout 0.05, AdamW with learning rate $2\times10^{-4}$, batch size 1, and 3 epochs \cite{dettmers2023qlora,loshchilov2019decoupled}. All experiments are run on a single NVIDIA RTX 4090 GPU. Unless otherwise stated, we set $w_{\mathrm{align}}=2.0$.

All methods use the same self-contained repair prompt and the same decoding configuration so that differences come from the training objective rather than prompt wording. For Defects4J, the prompt includes the title, natural-language bug description, filename, and buggy Java code. HumanEval-Java contains the filename and the buggy function. In both benchmarks, the buggy code is wrapped by the literal markers \texttt{<BOF>} and \texttt{<EOF>} and rendered inside a fenced Java block. We append an opening \texttt{```java} fence after \texttt{<EOF>} and extract the first fenced Java block as the candidate patch. The \prompt baseline differs only by inserting the phrase \emph{with the minimal change} into the final request sentence.

Backbone-specific chat wrappers are used only for serialization and share the same parsing logic for buggy and fixed code. At inference time, sampling uses a temperature of 0.7, a maximum generation length of 1024 tokens, and identical decoding termination criteria across all settings.

\subsection{Evaluation Metrics}
\label{sec:metrics}

We report functional correctness as pass@$k$ for $k \in \{1,5,10\}$. A candidate patch is considered plausible if it compiles and passes the benchmark test suite \cite{chen2021evaluatinglargelanguagemodels}.

To measure patch minimality, we compare each plausible predicted patch $P_i$ against its buggy input $B_i$ after applying a shared normalization function $\mathcal{N}(\cdot)$, which normalizes whitespace and removes comments. Let $\ell_i^X = \mathrm{Lines}(\mathcal{N}(X_i))$ and $t_i^X = T(\mathcal{N}(X_i))$ denote the normalized line and token sequences for $X \in \{B,P\}$. Unless otherwise noted, all minimality metrics are averaged over the subset of tasks for which a method produces at least one plausible patch.

We use Average Edit Distance (AED) as the primary edit-locality metric:
\begin{equation}
\mathrm{AED}
=
\frac{1}{N}\sum_{i=1}^{N}\operatorname{Lev}_{\mathrm{char}}\!\left(\mathcal{N}(B_i),\,\mathcal{N}(P_i)\right),
\end{equation}
where a lower AED indicates fewer character-level edits.

We use Code Consistency Rate (CCR) as a complementary preservation metric:
\begin{equation}
\mathrm{CCR}
=
\frac{1}{N}\sum_{i=1}^{N}\frac{k_i}{m_i},
\end{equation}
where $m_i = |\ell_i^P|$ and $k_i$ is the number of lines in $\ell_i^P$ that also appear verbatim in $\ell_i^B$ after normalization. Higher CCR indicates stronger preservation of stable context.

We additionally report Average Total Changed Lines (ATCL) and Average Total Changed Tokens (ATCT), defined as the total number of added and deleted lines or tokens in the normalized diff between the buggy code and the plausible patch.
\section{Experiments \& Results}
\label{sec:experiments}
\label{experiment}

\subsection{RQ1: Minimal-Edit Behavior and Metric Validity}
\label{sec:rq1}
\label{sec:empirical_min_edit}
\label{sec:human_annotation}

\noindent\textbf{[Experiment Goal for RQ1]:} We aim to understand whether current off-the-shelf repair models already produce minimal and review-friendly patches, and whether the automatic metrics used in this paper reflect human judgments of patch reasonableness. This RQ establishes both the need for preservation-aware training and the metric choice used in the later experiments.

\noindent\textbf{[Experiment Design for RQ1]:} We use two complementary analyses. First, we evaluate selected off-the-shelf code LLMs on Defects4J under the same decoding and evaluation pipeline, and report pass@$k$, CCR, AED, ATCL, and ATCT. Second, we conduct a human preference study on 50 randomly selected Defects4J bugs. For each bug, we present the buggy code and 2 plausible candidate patches in randomized order to 3 annotators, who choose the more reasonable patch based on semantic correctness and the extent of unnecessary modification. The study contains 2 rounds. Round I uses randomly sampled patch pairs. Round II restricts attention to cases where a metric-preferred patch conflicts with at least 1 size-related indicator. We measure inter-annotator reliability with Fleiss' $\kappa$, then compare metric-induced rankings with each annotator’s preference using Macro-F1 and Cohen's $\kappa$.

The annotation protocol was intentionally comparative. Each item contained the buggy snippet and two plausible patches with model identities removed. The left-right order of the two patches was randomized independently for each annotator. We recruited three annotators with Ph.D. degrees in computer science. Annotators were instructed to prefer the patch that best preserves the original intent while avoiding unnecessary modifications, assuming that both candidates pass the available tests. Ties were not allowed, so that every item yielded a forced preference. Tables~\ref{tab:metric_vs_human_round1}and~\ref{tab:metric_vs_human_round2} report per-annotator agreement, where metric-induced rankings are compared against each individual annotator’s preference. Because the conflict subset in Round II is defined separately for each metric, the evaluated patch pairs can differ across AED, CCR, ATCL, and ATCT; Table~\ref{tab:fleiss_kappa} therefore reports Fleiss' $\kappa$ on the metric-specific subsets used in each round.

\noindent\textbf{[Experimental Results for RQ1]:} Table~\ref{tab:defects4j-noft} shows that stronger off-the-shelf models achieve higher pass@1, but they still produce patches with substantial edit overhead. For example, Qwen3-Max and DeepSeek-V3 reach pass@1 of 30.4\% and 40.6\%, yet their AED remains 104.3 and 158.3, respectively. Models with similar CCR can also differ sharply in AED. This pattern shows that functional correctness and edit locality are not tightly coupled. Higher pass@$k$ does not guarantee a smaller or better-localized patch, which directly motivates a training objective that explicitly rewards the preservation of stable context.

\begin{table*}[h]
\centering
\caption{Performance of off-the-shelf code models on Defects4J. Best results in each column are shown in bold.}
\begin{tabular}{lccccccc}
\toprule
Model & pass@1 & pass@5 & pass@10 & CCR $\uparrow$ & AED $\downarrow$ & ATCL $\downarrow$ & ATCT $\downarrow$ \\
\midrule
\multicolumn{8}{c}{\textit{Open-source models}} \\

Qwen3-8B & 9.7 & 19.2 & 23.5 & 64.3 & 190.7 & 9.7 & 92.1 \\
OpenCoder-8B & 7.9 & 18.7 & 25.1 & \textbf{76.6} & 137.9 & 8.3 & 80.1 \\
DS-Coder-6.7B & 5.8 & 14.6 & 19.4 & 70.7 & 142.9 & 8.3 & 81.7 \\
\midrule
\multicolumn{8}{c}{\textit{Closed-source models}} \\

Qwen3-Max & 30.4 & 37.4 & 39.0 & 78.1 & \textbf{104.3} & 6.8 & 68.4 \\
DeepSeek-V3 & \textbf{40.6} & \textbf{48.8} & \textbf{51.1} & 76.3 & 158.3 & \textbf{6.7} & \textbf{62.2} \\
\bottomrule
\end{tabular}
\label{tab:defects4j-noft}
\end{table*}

Table~\ref{tab:fleiss_kappa} shows substantial and consistent agreement among annotators in both rounds, with Fleiss' $\kappa$ between 0.710 and 0.733. Tables~\ref{tab:metric_vs_human_round1} and~\ref{tab:metric_vs_human_round2} further show that AED and CCR align much better with human judgments than size-only metrics. In Round I, AED and CCR both achieve strong agreement across all annotators. In the conflict subset of Round II, AED remains the most reliable signal, with F1 of 0.929/0.857/0.852 and $\kappa$ of 0.838/0.676/0.678, while CCR degrades more noticeably and ATCL/ATCT fall to near-random or even negative agreement. We therefore use AED as the primary metric for patch quality and CCR as a secondary metric for stable-context preservation.

\begin{table}[H]
\centering
\caption{Fleiss' $\kappa$ for the three annotators. In Round II, the evaluated subset is metric-specific because conflict cases are defined by the metric-induced ranking.}
\label{tab:fleiss_kappa}
\begin{tabular}{lcc}
\hline
Metric & Round I & Round II \\
\hline
AED  & 0.733 & 0.729 \\
CCR  & 0.732 & 0.733 \\
ATCL & 0.733 & 0.710 \\
ATCT & 0.732 & 0.727 \\
\hline
\end{tabular}
\end{table}

\begin{table}[H]
\centering
\caption{Alignment between automatic metrics and human preferences in Round I.}
\label{tab:metric_vs_human_round1}
\begin{tabular}{lcccccc}
\hline
 & \multicolumn{6}{c}{Human Annotator} \\
Metric
& \multicolumn{2}{c}{Annotator 1}
& \multicolumn{2}{c}{Annotator 2}
& \multicolumn{2}{c}{Annotator 3} \\
 & F1 & $\kappa$ & F1 & $\kappa$ & F1 & $\kappa$ \\
\hline
AED  & 0.951 & 0.874 & 0.881 & 0.714 & 0.836 & 0.648 \\
CCR  & 0.963 & 0.919 & 0.893 & 0.757 & 0.846 & 0.680 \\
ATCL & 0.786 & 0.513 & 0.692 & 0.363 & 0.704 & 0.357 \\
ATCT & 0.724 & 0.343 & 0.643 & 0.190 & 0.667 & 0.280 \\
\hline
\end{tabular}
\end{table}

\begin{table}[H]
\centering
\setlength{\tabcolsep}{4pt}
\caption{Alignment between automatic metrics and human preferences in Round II (conflict subset).}
\label{tab:metric_vs_human_round2}
\begin{tabular}{lcccccc}
\hline
 & \multicolumn{6}{c}{Human Annotator} \\
Metric
& \multicolumn{2}{c}{Annotator 1}
& \multicolumn{2}{c}{Annotator 2}
& \multicolumn{2}{c}{Annotator 3} \\
 & F1 & $\kappa$ & F1 & $\kappa$ & F1 & $\kappa$ \\
\hline
AED  & 0.929 & 0.838 & 0.857 & 0.676 & 0.852 & 0.678 \\
CCR  & 0.826 & 0.680 & 0.773 & 0.595 & 0.739 & 0.520 \\
ATCL & 0.531 & 0.074 & 0.533 & 0.131 & 0.453 & $-0.137$ \\
ATCT & 0.500 & $-0.016$ & 0.480 & $-0.025$ & 0.520 & 0.054 \\
\hline
\end{tabular}
\end{table}

\find{\textbf{[RQ-1] Findings:} Even the strongest off-the-shelf models still over-edit, with DeepSeek-V3 reaching 40.6\% pass@1 but AED still at 158.3; in the human study, AED is the most reliable metric on the conflict subset, with F1 up to 0.929 and $\kappa$ up to 0.838. \textbf{Insights:} The main quality risk in LLM-based repair is often not a missed fix but a misplaced diff: a model can repair the observed behavior while editing well beyond the causal fault, and AED exposes that error earlier than pass@k.}

\subsection{RQ2: Effectiveness of \tool across Benchmarks}
\label{sec:rq2}

\noindent\textbf{[Experiment Goal for RQ2]:} We aim to evaluate whether preser\-vation-aware supervision improves both repair correctness and patch maintainability across benchmarks and backbone models.

\noindent\textbf{[Experiment Design for RQ2]:} For each backbone, we compare 3 settings: (i) \baseline, the off-the-shelf model; (ii) \stdft, standard supervised fine-tuning with the same data and QLoRA adapters but without preservation weighting; and (iii) \tool, our alignment-weighted fine-tuning method. We evaluate Qwen3-8B, OpenCoder-8B, and DS-Coder-6.7B on Defects4J and HumanEval-Java using the same prompt templates, decoding configuration, and evaluation harness. AED and CCR are computed on plausible patches and averaged over the subset of tasks with plausible outputs. Loss masking details are described in Section~\ref{sec:approach} and are further isolated in RQ4.

\noindent\textbf{[Experimental Results for RQ2]:} On Defects4J, Table~\ref{tab:defects4j-tool} shows that \tool improves pass@1 for all 3 backbones and substantially reduces AED relative to both \baseline and \stdft. The strongest gain appears on DS-Coder-6.7B, where \tool raises pass@1 from 5.8\% to 10.1\%, reduces AED from 142.9 to 80.7, and increases CCR from 70.7 to 76.3. Qwen3-8B and OpenCoder-8B show the same pattern. On HumanEval-Java, Table~\ref{tab:humanevaljava-tool} shows that \tool again improves pass@1 for all 3 backbones and usually improves AED and CCR over \stdft, with the largest pass@1 gain on OpenCoder-8B from 33.0\% to 50.4\%. Although CCR is not uniformly improved in every setting, the overlap analysis shows that \tool retains a large shared core with \baseline and \stdft while contributing the largest unique region, including 35 unique instances on Defects4J and 20 on HumanEval-Java. These gains indicate that \tool does not merely shrink diffs; it learns to preserve stable spans while still improving repair success. Compared to \baseline and \stdft, \tool contributes the largest unique solved set on Defects4J, with 35 instances solved only by \tool, compared with 12 for \baseline and 7 for \stdft.

\begin{table*}[t]
\centering
\caption{Effectiveness of \tool on Defects4J.}
\label{tab:defects4j-tool}
\begin{tabular}{l|l|ccccc}
\hline
Model & Method & @1 & @5 & @10 & AED $\downarrow$ & CCR $\uparrow$ \\
\hline
\multirow{3}{*}{Qwen3-8B}
& \baseline & 9.7  & 19.2 & 23.5 & 190.7 & 64.3 \\
& \stdft    & 10.2 {\up{5.2}}  & 23.0 {\up{19.8}} & 30.0 {\up{27.7}} & 101.8 {\up{46.6}} & 72.9 {\up{13.4}} \\
& \tool     & 13.0 {\up{34.0}} & 25.3 {\up{31.8}} & 30.5 {\up{29.8}} & 68.6 {\up{64.0}}  & 76.4 {\up{18.8}} \\
\hline
\multirow{3}{*}{OpenCoder-8B}
& \baseline & 7.9  & 18.7 & 25.1 & 137.9 & 76.6 \\
& \stdft    & 6.8 {\down{13.9}} & 17.3 {\down{7.5}} & 22.9 {\down{8.8}} & 110.2 {\up{20.1}} & 76.7 {\up{0.1}} \\
& \tool     & 10.1 {\up{27.8}}  & 22.5 {\up{20.3}}  & 28.3 {\up{12.7}} & 78.8 {\up{42.9}}  & 78.1 {\up{2.0}} \\
\hline
\multirow{3}{*}{DS-Coder-6.7B}
& \baseline & 5.8  & 14.6 & 19.4 & 142.9 & 70.7 \\
& \stdft    & 6.1 {\up{5.2}} & 15.1 {\up{3.4}} & 18.9 {\down{2.6}} & 93.5 {\up{34.6}}  & 70.6 {\down{0.1}} \\
& \tool     & 10.1 {\up{74.1}} & 21.2 {\up{45.2}} & 26.4 {\up{36.1}} & 80.7 {\up{43.5}}  & 76.3 {\up{7.9}} \\
\hline
\end{tabular}
\end{table*}

On HumanEval-Java, Table~\ref{tab:humanevaljava-tool} shows that \tool generalizes beyond project-level repair and continues to improve correctness under a function-level setting. Relative to \baseline, \tool sharply improves pass@1 and reduces AED for all 3 backbones. Relative to \stdft, \tool consistently improves pass@1 and usually improves AED and CCR. The largest gain appears on OpenCoder-8B, where pass@1 increases from 33.0\% under \stdft to 50.4\% under \tool, while AED drops from 100.4 to 74.2 and CCR rises from 69.9 to 74.8. Qwen3-8B also improves from 51.3\% to 57.3\% in pass@1 while reducing AED from 76.0 to 69.2. The main exception is DS-Coder-6.7B, where \tool improves pass@10 from 71.8\% to 78.3\% with better AED, but CCR decreases from 71.4 to 67.2. This exception confirms that correctness gains do not automatically imply better preservation, which is why both AED and CCR should be reported. Overlap analysis shows the same overall trend at the instance level: \tool contributes the largest unique solved set on HumanEval-Java, with 20 instances solved only by \tool, compared with 4 for \stdft and 0 for \baseline.

\begin{table*}[t]
\centering
\setlength{\tabcolsep}{4pt}
\caption{Effectiveness of \tool on HumanEval-Java.}
\label{tab:humanevaljava-tool}
\begin{tabular}{l|l|ccccc}
\hline
Model & Method & @1 & @5 & @10 & AED $\downarrow$ & CCR $\uparrow$ \\
\hline
\multirow{3}{*}{Qwen3-8B}
& \baseline & 44.8 & 69.6 & 76.5 & 257.7 & 51.3 \\
& \stdft    & 51.3 {\up{14.5}} & 77.9 {\up{11.9}} & 83.2 {\up{8.8}} & 76.0 {\up{70.5}}  & 74.3 {\up{44.8}} \\
& \tool     & 57.3 {\up{27.9}} & 82.1 {\up{18.0}} & 86.8 {\up{13.5}} & 69.2 {\up{73.1}}  & 74.6 {\up{45.4}} \\
\hline
\multirow{3}{*}{OpenCoder-8B}
& \baseline & 30.9 & 72.4 & 85.1 & 233.3 & 55.3 \\
& \stdft    & 33.0 {\up{6.8}} & 72.3 {\down{0.1}} & 82.6 {\down{2.9}} & 100.4 {\up{57.0}} & 69.9 {\up{26.4}} \\
& \tool     & 50.4 {\up{63.1}} & 79.6 {\up{9.9}}  & 86.3 {\up{1.4}}  & 74.2 {\up{68.2}}  & 74.8 {\up{35.3}} \\
\hline
\multirow{3}{*}{DS-Coder-6.7B}
& \baseline & 11.0 & 35.7 & 49.7 & 285.2 & 56.1 \\
& \stdft    & 19.1 {\up{73.6}} & 55.5 {\up{55.5}} & 71.8 {\up{44.5}} & 112.8 {\up{60.4}} & 71.4 {\up{27.3}} \\
& \tool     & 21.7 {\up{97.3}} & 60.8 {\up{70.3}} & 78.3 {\up{57.5}} & 103.3 {\up{63.8}} & 67.2 {\up{19.8}} \\
\hline
\end{tabular}
\end{table*}

\begin{table*}[t]
\centering
\setlength{\tabcolsep}{4pt}
\caption{Comparison on DS-Coder-6.7B for Defects4J. Min@10 (resp. Max@10) selects the plausible patch with the smallest AED (resp. the largest CCR) among the $k{=}10$ candidates.}
\label{tab:repairllama-compare}
\adjustbox{max width=\linewidth}{
\begin{tabular}{l|ccccc}
\hline
Method & pass@1 & pass@5 & pass@10 & AED (Avg./Med./Min@10)$\downarrow$ & CCR (Avg./Med./Max@10)$\uparrow$ \\
\hline
\baseline
& 5.8 
& 14.6 
& 19.4 
& 142.9  / 113.0  / 95.5 
& 70.7  / 77.0  / 77.5  \\

\stdft
& 6.1 {\up{5.2}}
& 15.1 {\up{3.4}}
& 18.9 {\down{2.6}}
& 93.5 {\up{34.6}} / 73.5 {\up{35.0}} / 54.7 {\up{42.7}}
& 70.6 {\down{0.1}} / 78.2 {\up{1.6}} / 80.7 {\up{4.1}} \\

\prompt
& 8.3 {\up{43.1}}
& 19.3 {\up{32.2}}
& 24.7 {\up{27.3}}
& 148.6 {\down{4.0}} / 104.5 {\up{7.5}} / 77.6 {\up{18.7}}
& 72.2 {\up{2.1}} / 75.0 {\down{2.6}} / 79.3 {\up{2.3}} \\

RepairLLaMA
& 7.0 {\up{20.7}}
& 17.7 {\up{21.2}}
& 23.7 {\up{22.2}}
& 149.1 {\down{4.3}} / 111.0 {\up{1.8}} / 105.4 {\down{10.4}}
& 35.3 {\down{50.1}} / 25.0 {\down{67.5}} / 56.8 {\down{26.7}} \\

AdaPatcher
& 5.9 {\up{1.7}}
& 16.6 {\up{13.7}}
& 22.1 {\up{13.9}}
& 112.4 {\up{21.3}} / 61.0 {\up{46.0}} / 77.0 {\up{19.4}}
& \textbf{79.3} {\up{12.2}} / \textbf{87.5} {\up{13.6}} / \textbf{84.2} {\up{8.6}} \\

\tool
& \textbf{10.1} {\up{74.1}}
& \textbf{21.2} {\up{45.2}}
& \textbf{26.4} {\up{36.1}}
& \textbf{80.7} {\up{43.5}} / \textbf{42.0} {\up{62.8}} / \textbf{48.9} {\up{48.8}}
& 76.3 {\up{7.9}} / 85.4 {\up{10.9}} / 81.8 {\up{5.5}} \\

\hline
\end{tabular}
}
\end{table*}

\find{\textbf{[RQ-2] Findings:} Across all settings, \tool improves pass@1 over \stdft, with gains up to 52.7\% on HumanEval-Java and up to 65.6\% relative on Defects4J, while reducing AED among all settings. \textbf{Insights:} \tool changes the repair distribution rather than merely shrinking outputs: it makes the model treat surrounding code as default context to preserve, so gains in correctness come from better edit placement rather than broader rewriting.}

\subsection{RQ3: Comparison with Representative Baselines}
\label{sec:rq3}

\noindent\textbf{[Experiment Goal for RQ3]:} We aim to determine whether the gains of \tool come from preservation-aware learning itself, rather than from a minimal-change prompt or from generic fine-tuning.

\noindent\textbf{[Experiment Design for RQ3]:} We fix DS-Coder-6.7B and evaluate on Defects4J. We compare \baseline, \stdft, \prompt, RepairLLaMA~\cite{silva2025repairllama}, AdaPatcher~\cite{dai2025less}, and \tool under the same evaluation harness and decoding configuration. The \prompt setting differs from \baseline only by adding an explicit instruction requesting minimal changes. We also report Min@10 AED and Max@10 CCR, obtained by selecting for each bug the plausible candidate with the smallest AED or the largest CCR among the 10 sampled candidates. These statistics provide an optimistic bound for test-time selection policies, while all main comparisons still use the default single-sample decoding.

\begin{figure}[t]
\centering
\includegraphics[width=\linewidth]{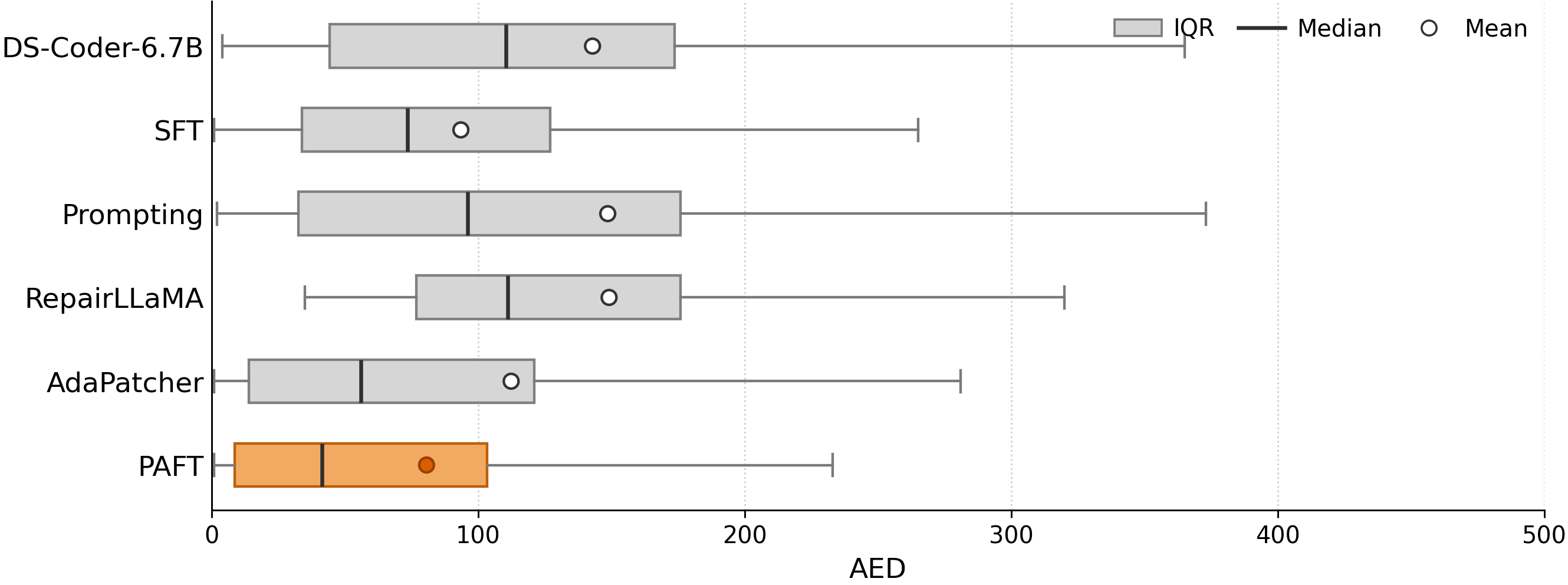}
\vspace{-0.6em}
\caption{Distribution of AED values over plausible patches on Defects4J for DS-Coder-6.7B and its variants. The first row corresponds to the base DS-Coder-6.7B model. \tool shifts the distribution toward smaller edits and achieves the lowest mean and median AED among all compared methods.}
\label{fig:rq3-aed-dist}
\end{figure}

\noindent\textbf{[Experimental Results for RQ3]:} Table~\ref{tab:repairllama-compare} shows that \tool achiev\-es the best pass@$k$ and the lowest AED across average, median, and Min@10 statistics. Figure~\ref{fig:rq3-aed-dist} provides a distributional view over plausible patches and shows the same pattern: \tool shifts the AED distribution leftward and attains the lowest mean and median AED among all methods. Compared with RepairLLaMA, \tool raises pass@1 from 7.0\% to 10.1\%, reduces average/median AED from 149.1/111.0 to 80.7/42.0, and increases average/median CCR from 35.3/25.0 to 76.3/85.4. Compared with \prompt, \tool still improves pass@1 from 8.3\% to 10.1\% while sharply reducing edits from 148.6/104.5 to 80.7/42.0 under average/median AED, showing that a prompt-only minimal-change instruction does not reliably control edit scope. Compared with AdaPatcher, \tool improves pass@1 from 5.9\% to 10.1\% and reduces AED from 112.4/61.0/77.0 to 80.7/42.0/48.9, although AdaPatcher remains higher on CCR. This difference matters because higher CCR alone does not yield better overall repair quality when plausibility is lower and character-level edits remain larger. Even with oracle selection among 10 samples, \baseline reaches AED Min@10 of 95.5 and CCR Max@10 of 77.5, both still worse than \tool's single-sample median AED and CCR of 42.0 and 85.4. Overall, \tool provides the strongest trade-off between correctness, locality, and preservation, and it consistently outperforms \stdft in both plausibility and edit locality.

\find{\textbf{[RQ-3] Findings:} On Defects4J with DS-Coder-6.7B, \tool achieves the best overall trade-off, raising pass@1 to 10.1\% versus 8.3\% for Prompting, 7.0\% for RepairLLaMA, and 5.9\% for AdaPatcher, while lowering average AED to 80.7 versus 148.6, 149.1, and 112.4. \textbf{Insights:} Outperforming strong baselines suggests that preservation-aware fine-tuning is an effective general recipe for tasks that require both accuracy and controlled modification, not only for minimal-edit program repair.}

\subsection{RQ4: Which Components of \tool Matter Most?}
\label{sec:rq4}

\noindent\textbf{[Experiment Goal for RQ4]:} We aim to isolate the contribution of each design choice in \tool, including standard fine-tuning, curriculum learning, alignment weighting, and full-sequence masking.

\noindent\textbf{[Experiment Design for RQ4]:} We fix DS-Coder-6.7B and evaluate all variants on Defects4J under the same data, training budget, and decoding protocol as in the main experiments. Starting from the base checkpoint, we progressively add (i) standard supervised fine-tuning on the target patch tokens, (ii) full-sequence masking, and (iii) an edit-difficulty curriculum. We then sweep the alignment weight \(w_{\mathrm{align}}\) while keeping the rest of the recipe fixed. Here, \(w_{\mathrm{align}}=1.0\) disables token upweighting, \(w_{\mathrm{align}}=2.0\) is the default setting, \(w_{\mathrm{align}}=0.0\) ignores the loss on aligned response tokens, and \(w_{\mathrm{align}}=4.0\) is an over-weighted diagnostic variant. We also compare variants without full masking, including an assistant-only masking configuration, to isolate the value of full-sequence masking.

\noindent\textbf{[Experimental Results for RQ4]:} Table~\ref{tab:loss-ablation} shows that standard fine-tuning mainly improves minimality rather than correctness. Relative to the base checkpoint, fine-tuning reduces AED from 142.9 to 93.5 while leaving pass@1 nearly unchanged at 6.1. Adding full-sequence masking improves pass@1 to 7.4 and CCR to 71.4, but the subsequent curriculum step pushes AED up to 111.2 and lowers CCR to 68.3, showing that diff-size ordering alone can bias training toward more aggressive edits when preservation is not explicitly optimized. Variants without full masking are consistently weaker than full \tool. For example, the assistant-only configuration with \(w_{\mathrm{align}}=2.0\) reaches pass@1/5/10 of 9.8/19.5/22.9 with AED 112.2 and CCR 75.1, which remains worse than full \tool on correctness and edit distance.

Sweeping \(w_{\mathrm{align}}\) shows that \(w_{\mathrm{align}}=2.0\) provides the best overall trade-off. Relative to the curriculum variant with \(w_{\mathrm{align}}=1.0\), the default setting improves pass@1 from 8.3 to 10.1, reduces AED from 111.2 to 80.7, and increases CCR from 68.3 to 76.3. Setting \(w_{\mathrm{align}}=0.0\) increases CCR to 75.8, but it hurts pass@1 and AED, indicating that preservation without sufficient learning on edit tokens is not enough. Increasing \(w_{\mathrm{align}}\) further to 4.0 lowers pass@1 to 8.2 and raises AED to 106.9 while only slightly improving CCR to 77.8. This pattern suggests that overly strong upweighting over-regularizes the model toward copying and weakens learning on the actual edit tokens.

\begin{table*}[t]
\centering
\small
\caption{Ablation with DS-Coder-6.7B on Defects4J.}
\label{tab:loss-ablation}
\setlength{\tabcolsep}{14pt}
\renewcommand{\arraystretch}{1.1}

\begin{tabular}{l ccccc}
\toprule
\textbf{Method} & \textbf{pass@1} & \textbf{pass@5} & \textbf{pass@10} & \textbf{AED} $\downarrow$ & \textbf{CCR} $\uparrow$ \\
\midrule
Base checkpoint & 5.8 & 14.6 & 19.4 & 142.9 & 70.7 \\

+ Fine-tuning
& 6.1 \up{5.2}
& 15.1 \up{3.4}
& 18.9 \down{2.6}
& 93.5 \up{34.6}
& 70.6 \down{0.1} \\

+ Full masking
& 7.4 \up{27.6}
& 18.8 \up{28.8}
& 25.6 \up{32.0}
& 105.0 \up{26.5}
& 71.4 \up{1.0} \\

+ Curriculum
& 8.3 \up{43.1}
& 19.9 \up{36.3}
& \textbf{26.4} \up{36.1}
& 111.2 \up{22.2}
& 68.3 \down{3.4} \\

\textbf{\tool (Best)}
& \textbf{10.1} \up{74.1}
& \textbf{21.2} \up{45.2}
& \textbf{26.4} \up{36.1}
& \textbf{80.7} \up{43.5}
& 76.3 \up{7.9} \\

\midrule
- Full masking
& 8.7 \up{50.0}
& 18.4 \up{26.0}
& 22.6 \up{16.5}
& 110.7 \up{22.5}
& 74.3 \up{5.1} \\

\tool (assistant-only, $w_{\mathrm{align}}=2.0$)
& 9.8 \up{69.0}
& 19.5 \up{33.6}
& 22.9 \up{18.0}
& 112.2 \up{21.5}
& 75.1 \up{6.2} \\

\midrule
\tool ($w_{\mathrm{align}}=0.0$)
& 7.3 \up{25.9}
& 19.3 \up{32.2}
& 25.9 \up{33.5}
& 120.2 \up{15.9}
& 75.8 \up{7.2} \\

\tool ($w_{\mathrm{align}}=1.0$)
& 8.3 \up{43.1}
& 19.9 \up{36.3}
& \textbf{26.4} \up{36.1}
& 111.2 \up{22.2}
& 68.3 \down{3.4} \\

\tool ($w_{\mathrm{align}}=2.0$)
& \textbf{10.1} \up{74.1}
& \textbf{21.2} \up{45.2}
& \textbf{26.4} \up{36.1}
& \textbf{80.7} \up{43.5}
& 76.3 \up{7.9} \\

\tool ($w_{\mathrm{align}}=4.0$)
& 8.2 \up{41.4}
& 19.6 \up{34.2}
& 25.1 \up{29.4}
& 106.9 \up{25.2}
& \textbf{77.8} \up{10.0} \\

\bottomrule
\end{tabular}
\end{table*}

\find{\textbf{[RQ-4] Findings:} On Defects4J with DS-Coder-6.7B, adding preservation-aware weighting at $w_{align}=2.0$ raises pass@1 from 8.3\% to 10.1\% and reduces AED from 111.2 to 80.7 over the same curriculum variant without weighting, showing that the main gain comes from preservation-aware supervision rather than generic fine-tuning alone. \textbf{Insights:} A promising application of \tool is to use it alongside capability-oriented adaptation methods, enabling models to improve task accuracy while preserving more stable content.}
\section{Discussion}
\label{sec:discussion}

Our results suggest that minimal-edit repair should be treated as a training objective rather than merely a decoding preference or a post hoc ranking criterion. By making stable-context preservation explicit during fine-tuning, \tool improves correctness while reducing unnecessary rewriting, which is especially valuable in review-oriented repair settings where a patch must be both test-passing and easy to inspect. The human preference results further suggest that AED and CCR are useful operational proxies for review burden among plausible patches, although they should still be interpreted as approximations rather than direct measures of maintainability.

The current findings should be interpreted within the scope of the evaluated setting. We study single-file Java repair on Defects4J (with file-level bug locations) and HumanEval-Java, so the reported gains should not yet be taken as evidence for multi-file, cross-language, or repository-scale repair. Moreover, \tool uses token-level alignment as a lightweight proxy for preservation. This design is efficient and effective, but it can under-credit semantically equivalent rewrites and does not represent the full space of acceptable fixes, since each training instance provides only one reference patch.

\tool should also be viewed as a developer aid rather than a fully autonomous repair system. Since correctness is judged by the available test suites, a plausible patch may still be semantically incorrect when the test suites are incomplete. This is especially important for safety-critical code, where additional validation beyond test passing remains necessary. Our human study was designed to minimize ethical risk: participants compared code snippets only, no personally identifying information was collected, and the displayed examples were screened for identifying or offensive content. The experiments relied on established public datasets, and the released artifacts are limited to scripts and generated results. In practice, \tool is best suited to developer-in-the-loop workflows, where its main benefit is to reduce review overhead by producing patches that are both more accurate and more conservative in edit scope.
\section{Threats to Validity}
\label{sec:threats}

\paragraph{\textbf{External validity.}}
To reduce the risk that the observed gains are specific to a single model or evaluation setup, we evaluate \tool across three backbone code LLMs and two complementary Java benchmarks under a unified training and evaluation protocol. These settings cover both function-level and project-level repairs, and the consistent improvements across them provide evidence that the method is not tied to any particular backbone or narrow repair scenario. The remaining external-validity threat concerns the breadth of transfer rather than whether the method works beyond a single setting.

\paragraph{\textbf{Internal validity.}}
A central internal-validity threat is whether the observed improvements can be attributed to preservation-aware supervision rather than to unrelated implementation or protocol differences. We reduce this threat by using the same training data, prompt structure, parsing rules, decoding configuration, and unified evaluation harness across Base, \stdft, and \tool, so the compared settings differ only in the intended training recipe. We further mitigate effect-attribution concerns by conducting ablations that separately examine standard fine-tuning, full-sequence masking, curriculum learning, and alignment weighting, all using the same backbone and training budget. These results show that the full \tool recipe is consistently stronger than the corresponding variants and that alignment weighting is a major source of the gain, even though interactions among components also contribute to the final outcome. We also reduce data-leakage risk through an exact SHA-256 contamination check between normalized training snippets and both evaluation benchmarks. As in all stochastic fine-tuning and sampling-based generation settings, run-to-run variation may affect absolute scores, but the consistency of the gains across models, benchmarks, and ablations reduces the likelihood that the main conclusions are artifacts of a particular run.

\paragraph{\textbf{Construct validity.}}
Our construct is intentionally narrow. We study review-oriented minimal-edit repair among plausible patches, rather than claiming to measure maintainability in full. Accordingly, pass@k together with benchmark compilation and test execution operationalizes repair plausibility, while AED, CCR, ATCL, and ATCT operationalize edit locality and stable-context preservation. These metrics do not cover every aspect of semantic equivalence, developer intent, or long-term maintainability, but they are well aligned with the concrete repair behavior targeted by \tool. We further strengthen this construct with a human-annotation study of 50 Defects4J bugs, which shows that AED and CCR more closely align with human preferences than size-only indicators, especially when metrics disagree.
\section{Related Work}

\subsection{Automated Program Repair}
Automated program repair (APR) has substantially reduced the manual effort required for defect fixing, and LLM-based approaches have become the dominant paradigm, such as ChatRepair~\cite{xia2023automated}, which adopts a procedural conversational framework that decomposes repair into multiple interaction steps. However, most existing LLM-based APR methods optimize for a single objective, namely generating patches that pass the available test suite, while overlooking patch maintainability. Although such patches may pass all the tests, they often introduce redundant code or unnecessarily disrupt the original code structure~\cite{Ehsani2026Where}. A study of 567 patches generated by Claude Code found that although 83.8\% of the patches were ultimately accepted, 41.5\% still required manual revision before being merged, with most revisions involving refactoring and style alignment~\cite{Chatlatanagulchai2025on}. These observations motivate our focus on patch minimization in addition to test-suite passing, to reduce unnecessary edits and ease downstream review.

\subsection{Minimal-Edit APR}
To encourage APR systems to generate minimal patches, existing methods mainly rely on prompt design to steer LLMs toward preserving the original code intent and avoiding unnecessary modifications. However, prompt-based methods depend heavily on a model's ability to consistently follow task-specific instructions. Their behavior is often unstable across different defect categories, and they usually require manually tailored prompts for different models~\cite{Tian2024Evaluating}. To address this issue, AdaPatcher~\cite{dai2025less} proposes a preference-based fine-tuning framework that constructs a large-scale preference dataset of approximately 52,000 program repair triplets and adopts a two-stage training pipeline, first \stdft and then DPO, to guide the model toward smaller edits. This design, however, incurs substantially higher training cost~\cite{zhou2023LIMA} and additional overhead for preference data construction. Unlike AdaPatcher, PAFT relies only on paired buggy/fixed examples and avoids the additional preference-data construction and DPO stage. In our experiments, this simpler setup yields higher pass rates and lower AED.

\section{Conclusion}
\tool addresses a specific weakness of LLM-based program repair: plausible patches often rewrite stable code outside the fault region. By making token-level preservation explicit during fine-tuning, \tool improves pass@1 over standard \stdft in all 6 evaluated backbone-benchmark settings on Defects4J and HumanEval-Java, with gains of up to 65.6\%, while reducing AED by up to 32.6\%. On Defects4J with DS-Coder-6.7B, \tool also achieves a better correctness-locality trade-off than representative baselines, reaching 10.1\% pass@1 with an AED of 80.7, compared with 8.3\% and 148.6 for Prompting, 7.0\% and 149.1 for RepairLLaMA, and 5.9\% and 112.4 for AdaPatcher. The annotation study further shows that AED is the strongest automatic proxy for review-oriented patch quality. These results suggest that preservation-aware strategies should be built into fine-tuning rather than left to prompting alone. \tool could be viewed as a complementary fine-tuning principle rather than a standalone recipe, where combined with other task-specific adaptation strategies to improve task accuracy while preserving stable context.




\balance
\bibliographystyle{ACM-Reference-Format}
\bibliography{sample-base}



\end{document}